\documentclass[aps,floatfix,showpacs,showkeys,amsmath]{revtex4}
\usepackage{color,graphicx}
\usepackage{dcolumn}
\usepackage{bm}
\usepackage{epsfig}
\usepackage{supertabular}
\usepackage[bookmarksopen]{hyperref}
\def\be {\begin{equation}}
\def\ee {\end{equation}}
\def\nn {\nonumber}
\def\bea {\begin{eqnarray}}
\def\eea {\end{eqnarray}}

\def  \p    {\pi}

\def  \f    {\frac}

\def  \veps {\varepsilon}

\def  \bef  {\begin{figure}}
\def  \eef  {\end{figure}}
\def  \be   {\begin{equation}}
\def  \ee   {\end{equation}}
\def  \ba   {\begin{array}}
\def  \ea   {\end{array}}
\def  \bea  {\begin{eqnarray}}
\def  \eea  {\end{eqnarray}}
\def  \beq  {\begin{eqnarray}}
\def  \eeq  {\end{eqnarray}}
\def  \nn   {\nonumber}
\def  \bd   {\begin{displaymath}}
\def  \ed   {\end{displaymath}}
\def  \bse  {\begin{subequations}}
\def  \ese  {\end{subequations}}
\def  \bwt  {\begin{widetext}}
\def  \ewt  {\end{widetext}}

\def  \ba   {{\bf{a_1}}}

\begin{document} 
 \title{Role of magnetic interactions in neutron stars}
\bigskip
\bigskip
\author{S.~P.~Adhya}
\email{souvikpriyam.adhya@saha.ac.in}
\author{ Pradip Roy}
\email{pradipk.roy@saha.ac.in}
\affiliation{
High Energy Nuclear and Particle Physics Division,
Saha Institute of Nuclear Physics, 1/AF Bidhannagar
Kolkata - 700064, India}

\begin{abstract}
In this work, we present a calculation of the non-Fermi liquid correction to the specific heat of magnetized degenerate quark matter present at the core of the neutron star. The role of non-Fermi liquid corrections to the neutrino emissivity has been calculated beyond leading order. We extend our result to the evaluation of the pulsar kick velocity and cooling of the star due to such anomalous corrections and present a comparison with the simple Fermi liquid case.
\end{abstract}
\maketitle

\section{Introduction}
\label{intro}
One of the fundamental aims of high energy physics is to study the properties of matter under extreme conditions i.e. in the domain of high temperature or high density where the hadrons are expected to melt into a plasma of quarks and gluons. Several efforts have been directed for the last two decades mainly to uncover the properties of this plasma at high temperature with zero chemical potential. In the present paper, we mainly focus on the other realm of phase space i.e. the region of dense ultrarelativistic plasma.

It has been recently shown that the behavior of quantum liquids in the ultrarelativistic regime is very different from the normal Fermi liquid (FL) behavior. This difference of behavior has recently been exposed both in the context of quantum electrodynamics (QED) and quantum chromodynamics (QCD)\cite{holstein73, bellac97}. 
It might be mentioned here that for the case of non-relativistic (NR) plasma, the magnetic interaction is supressed in powers of $(v/c)^2$ and hence can be neglected. For the case of NR plasmas; due to the absence of magnetic screening, Fermi liquid (FL) is sufficient to describe all the phenomenon. However for the case of relativistic degenerate plasmas, including the magnetic interactions change the FL picture significantly. Thus it has been seen that for the case of relativistic 
ultradegenerate plasma, in the vicinity of the Fermi surface, unscreened magnetostatic interactions lead to a logarithmic singularity in the inverse group velocity where the FL picture is no longer applicable \cite{rebhan05}. This non-Fermi liquid (NFL) behavior is manifested in the expression of mean free path (MFP), emissivity of neutrinos and specific heat of degenerate quark matter\cite{pal11,schafer04,holstein73, adhya12}.

Motivated with these results, we have recently derived the expressions of the neutrino mean free path (MFP) and the emissivity of the neutrinos from neutron star\cite{adhya12}. In our work, we have incorporated NFL corrections upto next to leading order (NLO) in degenerate quark matter that might exist in the core of the neutron star. All these calculations were previously restricted to the leading order (LO) containing the anomalous $T{ln}(1/T)$ term in these expressions\cite{pal11,schafer04}. We have extended our calculation beyond the known leading logarithmic order and found the appearance of fractional powers in $(T/\mu)$ (where T is the temperature and $\mu$ is the chemical potential of the the degenerate quark matter) in the expressions of MFP and emissivity. As an application, we then show how such corrections affect the neutrino emissivity for which we consider quark direct URCA process and inverse URCA processes given by \cite{iwamoto81},
\bea
&&d\rightarrow u+e^-+\bar{\nu_{e}}
\label{dir}
\eea
\bea
&&u+e^-\rightarrow d+\nu_{e}.
\label{inv}
\eea
 Subsequently, we study the cooling behavior of the neutron star (NS).
Exploration of the bizarre phenomenon of
pulsar kicks i.e. the observed large escape velocities of NS out of supernova
remnants has drawn significant attention in
recent years. For the past few years, it has
been argued that asymmetric neutrino emission is responsible for
the pulsar kicks during the evolution of the
 NS \cite{dorofeev85,sagertarxiv1,sagert08}. 
The URCA reactions, in presence of a strong magnetic field can give rise to asymmetric neutrino emission as explained in \cite{dorofeev85,sagert08}. Magnetic fields of the order of $10^{15}-10^{19} Gauss$ are known to exist in the core of the NS. Thus the electrons are forced to occupy the lowest Landau level polarising the electron spin opposite to the direction of the magnetic field. 
The electron polarisation for different conditions of magnetic field and kick velocities has been studied recently by Sagert et. al.\cite{sagert08}. In this work, the authors have studied pulsar acceleration mechanism based on asymmetric neutrino emission from quark direct URCA process.
Extension of our calculation of the modified dispersion relation also leads to the modification of the pulsar kick velocity of the neutron star. Additional corrections are included for the effect of external magnetic field on the specific heat capacity of the quarks which leads to further modification of the velocity.\\

In this work, we investigate non-Fermi liquid(NFL) behavior at NLO that enters into the
emissivity of the neutrinos from the NS composed of degenerate
quark matter core. Such calculation was first done in Ref.\cite{iwamoto81}
where the author studied only the Fermi liquid(FL) case. Recently, the
calculation was extended to the leading order (LO) NFL effect in
Ref.\cite{schafer04} and NLO in Ref.\cite{adhya12}.
It is in this context, we visit the problem of the calculation of the kick velocity to see whether such NFL corrections are significant to alter the kick velocity compared to the FL results. In addition, we study the effect of the external magnetic field in the kick velocity of the NS\cite{adhya14}.\\
The ppaer is organised as follows. In section II, we devolop the formalism by calculation of emissivity of neutrinos followed by deriving the expressions of the kick velocity of the NS. We discuss our results in Section III and summarizing in Section IV.
\section{Formalism}
\subsection*{Emissivity of neutrinos}

Now, the emissivity of the neutrinos is given by\cite{iwamoto81},
\bea
\veps=\int \frac{d^{3}p_{\nu}}{(2\pi)^{3}}E_{\nu}\frac{1}{l(-E_{\nu},T)}.
\eea
where, the mean free path of the neutrinos is given by\cite{iwamoto81},
\bea
\label{mfp01}
\frac{1}{l_{mean}^{abs}(E_{\nu},T)}=&&\frac{g^{\prime}}{2E_{\nu}}\int\frac{d^3p_d}{
(2\p)^3}
\frac{1}{2E_d}\int\frac{d^3p_u}{(2\p)^3}
\frac{1}{2E_u}\int\frac{d^3p_e}{
(2\p)^3 }
\frac{1}{2E_e}
(2\pi)^4\delta^4(P_d +P_{\nu}-P_u
-P_e)\nn\\
&&\times|M|^2 \{n(p_d)[1-n(p_u)][1-n(p_e)]
-n(p_u)n(p_e)[1-n(p_d)]\},
\eea
The Fermi liquid (FL) contribution has been calculated in the works of \cite{iwamoto81}.
We know that quasiparticle interactions is responsible for the modification of the dispersion relation given by\cite{bellac97},
\bea
\omega=(E_{p(\omega)} +{\rm Re}\Sigma(\omega,p(\omega)))
\eea
We mainly focus on the quasiparticle self energy of the quarks in the region of low temperature (T) and high chemical potential ($\mu$) for the calculation of associated quantities. The real part of the quasiparticle self energy ${\rm Re}\Sigma$ is given as \cite{rebhan05}:
\begin{eqnarray}
&&\rm{Re}\Sigma_{+}(\omega)=-g^2C_Fm\,
 \Big\{{\epsilon\over12\pi^2m}\Big[\log\Big({4\sqrt{2}m\over\pi
\epsilon}\Big)+1\Big]+{2^{1/3}\sqrt{3}\over45\pi^{7/3}}\left({\epsilon\over m}\right)^{5/3}
-20{2^{2/3}\sqrt{3}\over189\pi^{11/3}}\left({\epsilon\over m}\right)^{7/3}\nn\\
&&-{6144-256\pi^2+36\pi^4-9\pi^6\over864\pi^6}\Big({\epsilon\over m} 
  \Big)^3\Big[\log\left({{0.928}\,m\over \epsilon}\right)\Big]
  +\mathcal{O}\Big(\left({\epsilon\over m}\right)^{11/3}\Big)\Big\}
\end{eqnarray}
where $\epsilon=(\omega-\mu)$.
Thus we obtain, at the leading order \cite{schafer04},
\bea
\varepsilon_{LO} \simeq \frac{457}{3780}G_{F}^{2}cos^{2}\theta_{c}C_{F}\alpha_{s}\mu_{e}T^{6}\frac{(g\mu)^2}{\pi^2}\text{ln}\Big(\frac{4g\mu}{\pi^{2}T}\Big)
\eea
which agrees with the result quoted in ref.\cite{schafer04}.
Now, we obtain the NLO contribution to the neutrino emissivity as\cite{adhya12},
\bea
\varepsilon_{NLO} \simeq \frac{457}{315}G_{F}^{2}cos^{2}\theta_{c}C_{F}\alpha_{s}\mu_{e}T^{6}\Big[c_{1}T^{2} + c_{2}T^{2/3}(g\mu)^{4/3} - c_{3}T^{4/3}(g\mu)^{2/3} - c_{4}T^{2}\text{ln}\Big(\frac{0.656g\mu}{\pi T}\Big)\Big]
\eea
where the constants are evaluated as,
\bea
c_1 = -0.0036\pi^{2};
c_2 = \frac{2^{2/3}}{9\sqrt{3}\pi^{5/3}};
c_3 = \frac{40\times2^{1/3}}{27\sqrt{3}\pi^{7/3}}
\eea
and
\bea
c_4 = \frac{6144-256\pi^{2}+36\pi^{4}-9\pi^{6}}{144\pi^{4}}.
\eea
Thus we obtain the expressions for the emissivity of the neutrinos at the leading and the next to leading order.
\subsection*{Kick velocity}
The pulsar acceleration can be written as a function of the emissivity of the neutrinos, radius of the the quark core. In addition, it also depends on the polarisation of the electron spin and the mass of the neutron star. Thus we obtain\cite{sagertarxiv1}, 
\begin{eqnarray}
 dv=\frac{\chi}{M_{NS}}\frac{4}{3}\pi R^{3}\epsilon dt
 \label{diffv}
\end{eqnarray}
 Now using the cooling equation, we can rewrite the equation in terms of specific heat of the quark matter core.
\begin{eqnarray}
 C_{v}dT=-\epsilon dt
\end{eqnarray}
In recent literatures, the calculation of specific heat of quark matter is \cite{holstein73, ipp04},
\begin{eqnarray}
C_{v}\Big|_{FL}&=&\frac{N_{c}N_{f}}{3}\mu_{q}^{2}T
\end{eqnarray}
Thus the Fermi liquid contribution to the pulsar kick velocity as reported in \cite{sagert08} can be recast into the following form,
\begin{eqnarray}
 v\Big|_{FL}&\simeq&\frac{8.3 N_{C}N_{f}}{3}\Big(\frac{\mu_{q}}{400MeV}\frac{T}{1MeV}\Big)^{2}\Big(\frac{R}{10km}\Big)^{3}\frac{1.4M_{\odot}}{M_{NS}}\chi\frac{
km}{s}
\label{vfl}
\end{eqnarray}
Now extending the calculation of the pulsar kick velocity by incorporating the effect of the non-Fermi liquid 
effects into the specific heat can be written by using the modified dispersion relation.

 Thus the specific heat of the degenerate quark matter up to NLO is given by \cite{holstein73,ipp04},
\bea 
\label{spec-heat}
 \label{finalcv}
 C_v\Big{|}_{total}=C_v\Big{|}_{FL}+C_v\Big{|}_{LO}+C_v\Big{|}_{NLO}
 \eea
 where,
 \bea
 C_v\Big{|}_{LO}=N_g{g_{eff}^2\mu_q^2 T\over36\pi^2}\left(\ln\left({4g_{eff}\mu_q\over\pi^2T}\right)+\gamma_E
   -{6\over\pi^2}\zeta^\prime(2)-3\right)
 \eea
 and
 \bea
 C_v\Big{|}_{NLO}&=&N_g\Big[-40{2^{2/3}\Gamma\left({8\over3}\right)\zeta\left({8\over3}\right)\over27\sqrt{3}\pi^{11/3}}
   T^{5/3}(g_{eff}\mu_q)^{4/3}
   +560{2^{1/3}\Gamma\left({10\over3}\right)\zeta\left({10\over3}\right)
   \over81\sqrt{3}\pi^{13/3}}T^{7/3}(g_{eff}\mu_q)^{2/3}\nn\\
   &&+{2048-256\pi^2-36\pi^4+3\pi^6\over180\pi^2}T^3
   \Big[\ln\left({g_{eff}\mu_q\over T}\right)+\bar c-{\frac{7}{12}}\Big]\Big]
 \eea
 where the coupling constant $g$ is related to $g_{eff}$ as,
 \begin{equation}\label{geffdef}
 g^2 = \frac{2\ g^{2}_{eff}}{N_{f}},
 \end{equation}
 and $C_v\Big{|}_{total}$ is the sum of the FL, LO and NLO contribution to the specific heat of the quark matter.
Thus we obtain the LO and NLO contributions to the kick velocity as\cite{adhya14},
\begin{eqnarray}
v\Big|_{LO}\simeq\frac{16.6 N_{C}N_{f}}{3}(C_F\alpha_s)\Big(\frac{\mu_{q}}{400MeV}\frac{T}{1MeV}\Big)^{2}\Big(\frac{R}{10km}\Big)^{3}\frac{1.4M_{\odot}}{M_{NS}}\chi\Big[c_1+c_2\ln\Big(\f{g\mu_q\sqrt{N_f}}{T}\Big)\Big]\frac{
km}{s}
\label{vfllo}
\end{eqnarray}
where $C_F=(N_c^2-1)/(2N_c)$ and the constants are $c_{1}=-0.13807$ and $c_{2}=0.0530516$.
\begin{eqnarray}
 v\Big|_{NLO}&\simeq&\frac{16.6 N_{C}N_{f}}{3}\Big(\frac{\mu_{q}}{400MeV}\frac{T}{1MeV}\Big)^{2}\Big(\frac{R}{10km}\Big)^{3}\frac{1.4M_{\odot}}{M_{NS}}\chi(C_{F}\alpha_{s})\nn\\
 &\times&\Big[a_1\Big(\frac{bT}{\mu_q}
\Big)^{2/3}+a_2\Big(\frac{bT}{\mu_q}\Big)^{4/3}
+ \Big[a_3+a_4\ln\Big(\frac
{\mu_q}{b T}\Big)\Big]\Big(\frac{bT}{\mu_q}\Big)^2\Big]\frac{km}{s}
\end{eqnarray}
where the constants are evaluated as,
\begin{eqnarray}
 a_1=-\frac{12\pi\times0.04386}{8};a_2=\frac{12\pi\times0.04613}{10}
;a_3=-2.4162;a_4=-0.4595
\end{eqnarray}
and
\begin{eqnarray}
 b=\frac{2\pi}{\sqrt{N_f}g}.
\end{eqnarray}
The net contribution to the pulsar kick velocity upto NLO is obtained by the sum of the
Fermi liquid result and the non-Fermi liquid correction upto NLO:
\bea
v\Big{|}_{total}=v\Big|_{FL}+v\Big|_{LO}+v\Big|_{NLO}
\label{vtotal}
\eea
As high magnetic field exists in the core of the neutron star, therefore
the modification of the specific heat capacity in presence of such high 
magnetic field should also be considered. Thus, we have computed the specific heat capacity
of the quark matter in presence of such high magnetic field as\cite{adhya14},
\bea
C_{v}\Big{|}_{FL}^{B}=\f{N_CN_fTm_q^2}{6}\Big(\f{B}{B_{cr}^q}\Big)
\label{cvFLB}
\eea
Incorporating the effect of the NFL behavior in the specific heat capacity, we obtain,
\bea
C_v\Big{|}_{LO}^B\simeq\Big(\f{N_CN_fC_f\alpha_s}{36\pi}\Big)m_q^2\Big(\f{B}{B_{cr}^q}\Big)T\Big[(-1+2\gamma_E)+2log\Big(\f{2m_B}{T}\Big)\Big]
\eea
The NLO contribution to the specific heat capacity is obtained as,
\bea
C_v\Big{|}_{NLO}^B\simeq&&\Big(\f{N_CN_f}{3}\Big)(C_f\alpha_s)\Big(m_q^2\f{B}{B_{cr}^q}\Big)T\Big[c_1\Big(\f{T}{m_B}\Big)^{2/3}\nn\\
&+&c_2\Big(\f{T}{m_B}\Big)^{4/3}+c_{3}\Big(\f{T}{m_B}\Big)^{2}\Big(c_{4}-\log\Big(\f{T}{m_B}\Big)\Big)\Big]
\eea
where the constants are \cite{adhya14},
\bea
c_1=-0.2752; c_2=0.2899; c_3=-0.5919; c_4=5.007.
\eea
The Debye mass ($m_B$) in the QCD case in presence of magnetic field is obtained as follows\cite{rebhan05},
\bea
m_B^{2}=\f{N_fg^2m_q^2}{4\pi^2}\Big(\f{B}{B_{cr}^q}\Big)
\eea
The pulsar kick velocity obtained taking into account the magnetic field effect on the specific heat capacity of the quarks reads as \cite{adhya14},
\bea
 v\Big|_{FL}^B &\simeq&\frac{4.15 N_{C}N_f}{3}\Big(\frac{\sqrt{m_q^2(B/B_{cr}^q)}}{400MeV}\frac{T}{1MeV}\Big)^{2}\Big(\frac{R}{10km}\Big)^{3}\frac{1.4M_{\odot}}{M_{NS}}\chi\frac{
km}{s}
\label{vBFL}
\eea
By implementing the anomalous NFL effect, we obtain the LO contribution to the kick velocity as\cite{adhya14},
\bea
v\Big|_{LO}^B &\simeq&\frac{8.8 N_{C}N_f}{3}(C_f\alpha_s)\Big(\frac{\sqrt{m_q^2(B/B_{cr}^q)}}{400MeV}\frac{T}{1MeV}\Big)^{2}\Big(\frac{R}{10km}\Big)^{3}\nn\\
&&\times\frac{1.4M_{\odot}}{M_{NS}}\Big[0.0635+0.05\log\Big(\f{m_B}{T}\Big)\Big]\chi\frac{
km}{s}
\label{vBLO}
\eea
We have extended our calculation beyond the LO in NFL correction.
The NLO correction to the kick velocity is calculated as \cite{adhya14},
\bea
v\Big|_{NLO}^B&\simeq&\frac{8.3 N_{C}N_{f}}{3}\Big(\f{B}{B_{cr}^q}\Big)\Big(\frac{m_{q}}{400MeV}\frac{T}{1MeV}\Big)^{2}\Big(\frac{R}{10km}\Big)^{3}\frac{1.4M_{\odot}}{M_{NS}}\nn\\
&&\times\chi(C_{F}\alpha_{s})\Big[a_1\Big(\frac{T}{m_B}
\Big)^{2/3}+a_2\Big(\frac{T}{m_B}\Big)^{4/3}\nn\\
&&+ \Big[a_3+a_4\ln\Big(\frac
{m_B}{T}\Big)\Big]\Big(\frac{T}{m_B}\Big)^2\Big]\frac{km}{s}
\label{vBNLO}
\eea
The constants are evaluated as,
\begin{eqnarray}
 a_1=-\frac{12\pi\times0.04386}{8};a_2=\frac{12\pi\times0.04613}{10}
;a_3=-2.4162;a_4=-0.4595
\end{eqnarray}
It is to be noted that the magnetic interaction which have a long range character 
leads to the anomalous $T^2\log T^{-1}$ term in the pulsar velocity. The effect of the 
electron polarisation fraction for different condition of the magnetic field is used 
for the estimation of the kick velocity of the NS which we present in the Results section.
\section{Results and discussions}
\label{sec-2}
\begin{figure}[h]
 \centering
 \includegraphics[width=7cm,clip]{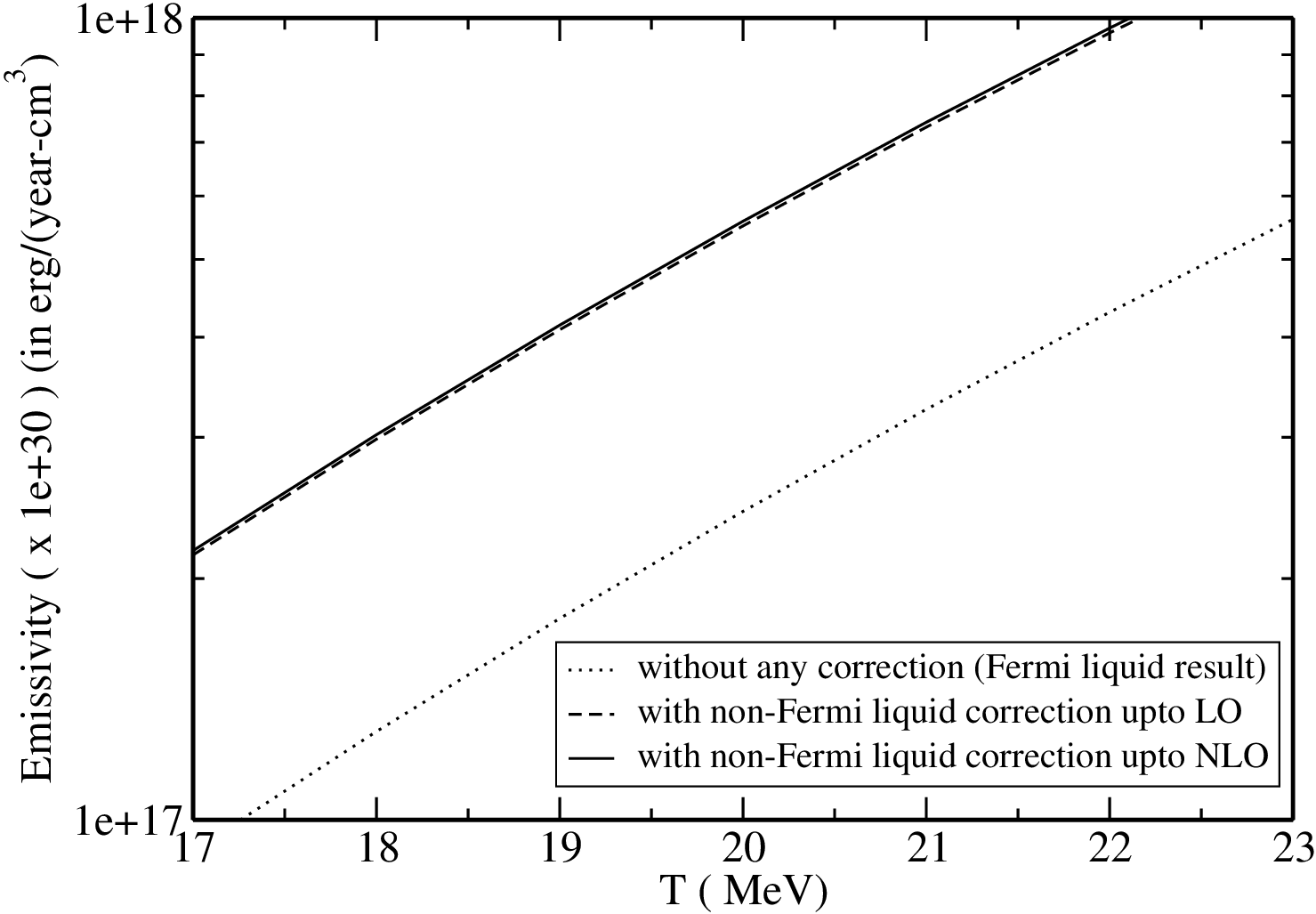}~~~~~~~~\includegraphics[width=7cm,clip]{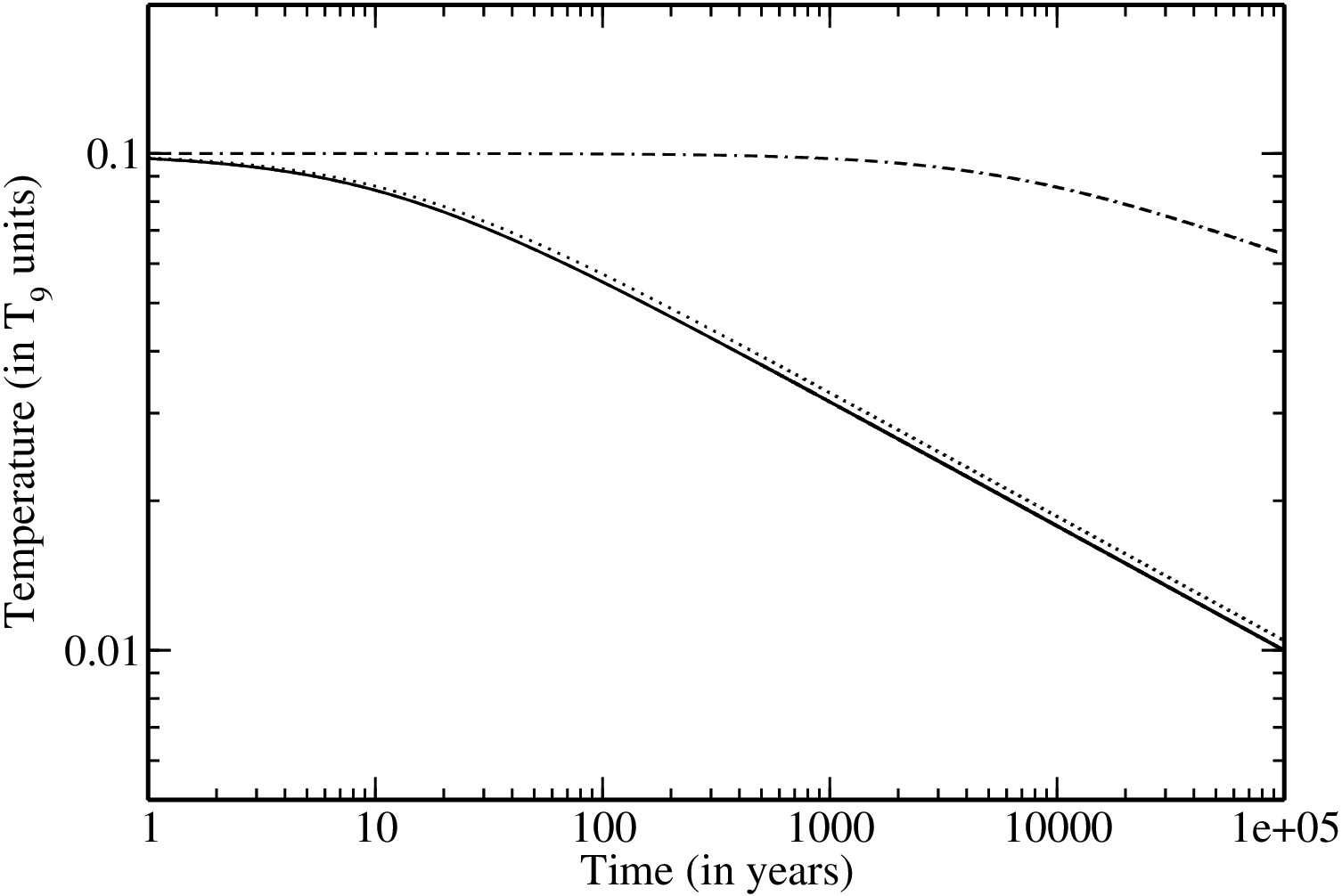}
 \caption{The left panel shows comparison of the emissivity of neutrinos for FL, LO and NLO NFL result with temperature of the quark matter core. The right panel shows the comparison between the cooling nature of the NS with neutron matter (dash dotted line) and quark matter. The dotted line
represents the FL contribution, the solid line represents the
NFL NLO correction.
}
 \label{fig1}       
 \end{figure}

\begin{figure}[h]
 \centering
 \includegraphics[width=7cm,clip]{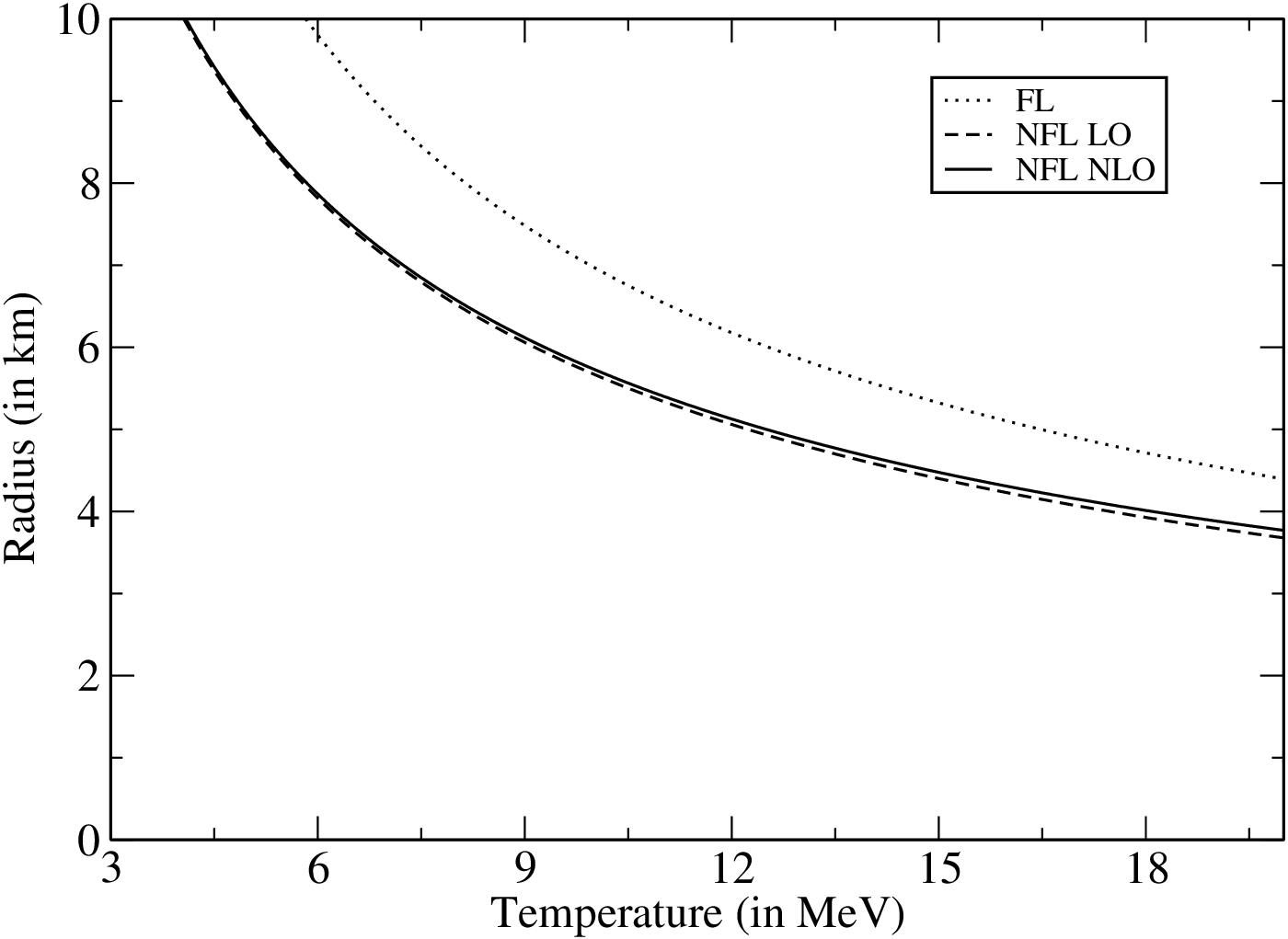}~~~~~~~~\includegraphics[width=7cm,clip]{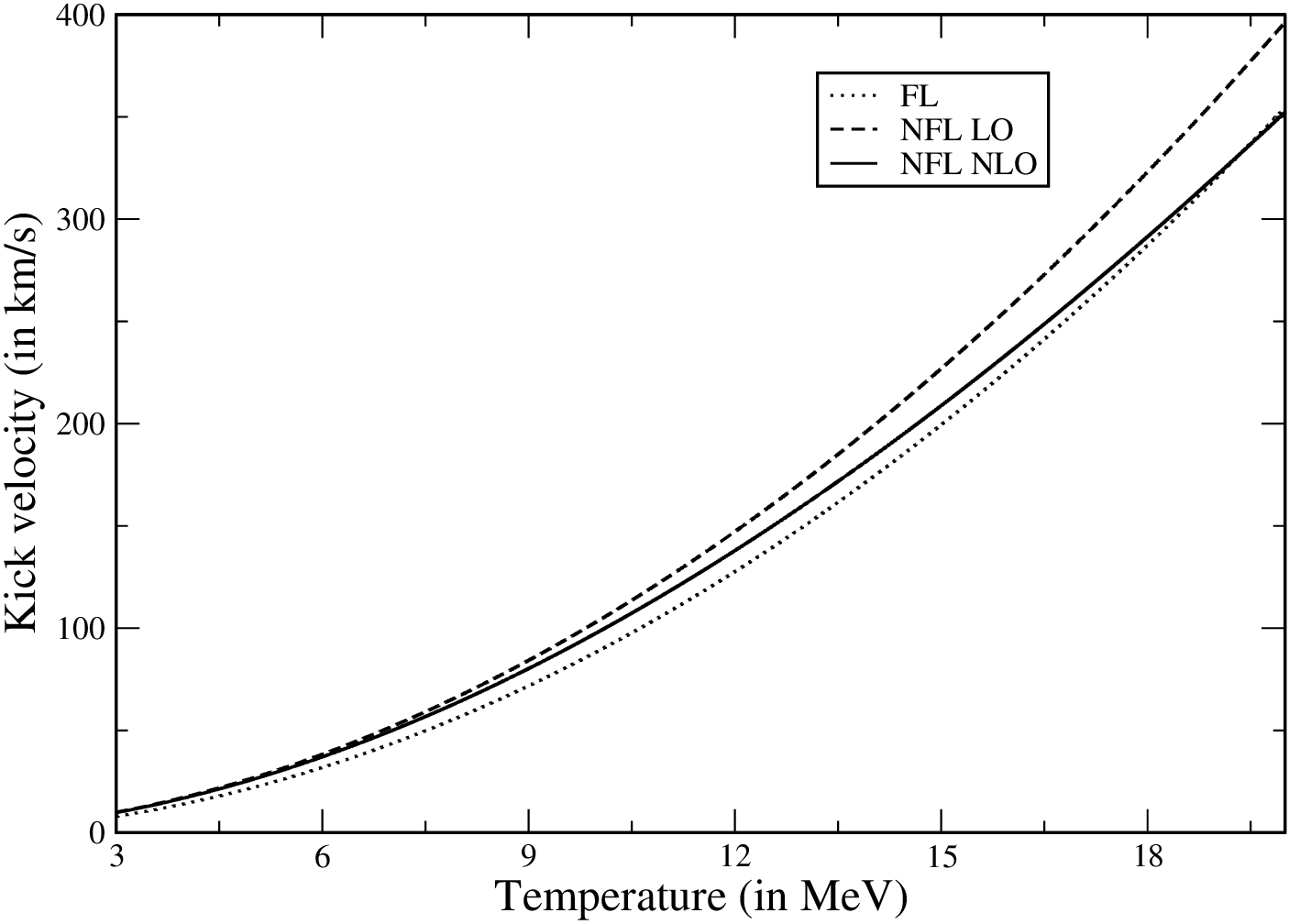}
 \caption{The left panel shows the comparison between radius and temperature of the core of NS for the case of high magnetic field. The right panel shows the comparison between kick velocity and temperature of the core for weak magnetic field. }
 \label{fig2}       
 \end{figure}

In this section assuming a quark chemical potential of $500 MeV$ and electron chemical potential of $15 MeV$, we have plotted the variation of the emissivity of the neutrinos with temperature present at the core of the neutron star. In the left panel of Fig.\ref{fig1} we show comparison of the neutrino emissivity with temperature for the FL, LO and NLO cases. The right panel shows a comparison of the cooling nature of the neutron star when the core is assumed to be QGP with that with normal nuclear matter.
In this section, we have also presented an estimation of the radius of the NS for different temperatures. The numerical computation of the kick velocity has been performed with different values of the polarisation fraction as in \cite{sagert08, sagertarxiv1} for weak and strong external magnetic fields. In the left panel of Fig.\ref{fig2}, we have plotted the quark matter core radius with the temperature considering the electrons are in highly polarised condition whereas in the right panel, we have expressed our results of the kick velocity with temperature assuming that the electrons are partially polarised. We note that the inclusion of the medium modified propagator increases the kick velocity substantially for the LO case. However, when we extend our results to the NLO case, we find that there is marginal change in the value of the kick velocity compared to the LO case.
\section{Summary and discussions}
The expression of emissivity of neutrinos by incorporating NFL effects upto next-to-leading order (NLO) is calculated. It is found that the emissivity contains terms at the higher order which involves fractional powers and logarithms in $(T/\mu)$. It is also found that there is increment of emissivity due to NLO corrections over FL and LO results. On examining the cooling behavior it is seen that the cooling is affected moderately compared to the simple FL case. 
 The results show that the pulsar kick velocity receives significant contribution from the logarithmic corrections. Further, incorporation of results upto next-to-leading order to include plasma/quasiparticle effects which are anomalous (NFL) effects has been done. The contribution from electron polarisation for different cases has been taken into account to calculate the velocities. The presence of the logarithmic term and the magnetic field considerably enhances the kick velocity of the neutron star.
\subsection*{Acknowledgements}
One of the authors [SPA] would like to thank UGC, India for providing the fellowship (Sr. no.: 2120951147) during the tenure of this work.

\end{document}